\documentclass{IEEEoj}
\usepackage{amsmath,amsfonts,amssymb,bm}
\usepackage{algorithmic}
\usepackage{algorithm}
\usepackage{booktabs}
\usepackage{array}
\usepackage{optidef}
\usepackage[caption=false]{subfig}

\DeclareMathOperator{\diag}{diag}
\usepackage{textcomp}
\usepackage{stfloats}
\usepackage{url}
\usepackage{verbatim}
\usepackage{graphicx}
\usepackage{cite}
\usepackage{siunitx}

\newcommand{\STATEnonum}{\item[]}

\def\BibTeX{{\rm B\kern-.05em{\sc i\kern-.025em b}\kern-.08em
    T\kern-.1667em\lower.7ex\hbox{E}\kern-.125emX}}
\AtBeginDocument{\definecolor{ojcolor}{cmyk}{0.93,0.59,0.15,0.02}}
\def\OJlogo{\vspace{-14pt}\includegraphics[draft=false,height=24pt]{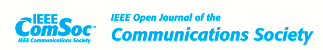}}
\begin{document}
\receiveddate{XX Month, XXXX}
\reviseddate{XX Month, XXXX}
\accepteddate{XX Month, XXXX}
\publisheddate{XX Month, XXXX}
\currentdate{XX Month, XXXX}
\doiinfo{OJCOMS.2022.1234567}

\title{Opportunistic Information-Bottleneck for Goal-oriented Feature Extraction and Communication}

\author{Francesco Binucci$^{1,2}$, Paolo Banelli$^1$, Paolo Di Lorenzo$^{2,3}$, Sergio Barbarossa$^3$
}
\affil{Department of Engineering, University of Perugia, Via G. Duranti 93, 06128, Perugia, Italy\vspace{.1cm}}
\affil{Consorzio Nazionale
Interuniversitario per le
Telecomunicazioni (CNIT), 43124, Parma, Italy}
\affil{Department of Information Engineering, 
Electronics, and Telecommunications, Sapienza University of Rome, via Eudossiana 18, 00184, Rome, Italy\vspace{.1cm}
}
\corresp{CORRESPONDING AUTHOR: Francesco Binucci (e-mail: francesco.binucci@studenti.unipg.it).}
\authornote{This work was supported by the European Union under the Italian National
Recovery and Resilience Plan (NRRP) of NextGenerationEU, partnership on
“Telecommunications of the Future” (PE00000001 - program “RESTART”), and by the 6G-GOALS project under
the 6G SNS-JU Horizon program, n.101139232. }

\begin{abstract}

The Information Bottleneck (IB) method is an information theoretical framework to design a parsimonious and tunable feature-extraction mechanism, such that the extracted features are maximally relevant to a specific learning or inference task. 
Despite its theoretical value, the IB is based on a functional optimization problem that admits a closed form solution only on specific cases (e.g., Gaussian distributions), making it difficult to be applied in most applications, where it is necessary to resort to complex and approximated variational implementations. 
To overcome this limitation, we propose an approach to adapt the closed-form solution of the Gaussian IB to a general task. Whichever is the inference task to be  performed by a (possibly deep) neural-network, the key idea is to \emph{opportunistically} design a regression sub-task, embedded in the original problem, where we can safely assume a (joint) multivariate normality between the sub-task's inputs and outputs. In this way we can exploit a fixed and pre-trained neural network to process the input data, using a tunable number of features, to trade data-size and complexity for accuracy.
This approach is particularly useful every time a device needs to transmit data (or features) to a server that has to fulfil an inference task, as it provides a principled way to extract the most relevant features for the task to be executed, while looking for the best trade-off between the size of the feature vector to be transmitted,  inference accuracy, and complexity.
Extensive simulation results testify the effectiveness of the proposed method and encourage to further investigate this research line.

\end{abstract}

\begin{IEEEkeywords}
Goal-Oriented communications, information bottleneck, edge-intelligence
\end{IEEEkeywords}

\maketitle

\section{Introduction}

Artificial Intelligence (AI) and Machine Learning (ML) are experiencing an unprecedented explosion nowadays, with a proliferation of applications and use-cases, such as Generative Deep Learning \cite{creswell2018generative}, Natural Language Processing (NLP) \cite{chowdhary2020natural}, object detection/recognition \cite{szegedy2013deep}, motion planning \cite{li2023edge}, and many others.
\\
The sixth generation of mobile networks is expected to be one of the main key-enablers for these applications \cite{strinati20196g}, with an ever-increasing integration of AI/ML blocks within the communication infrastructure. \textcolor{black}{In many ML application scenarios, there is the need to bring intelligence to peripheral devices (sensors), which however have very limited resources (e.g., IoT devices, Unnamed Aerial Vehicles (UAVs) \cite{kurunathan2023machine}, etc.). The problem is even more critical when the applications to be run are delay-sensitive, or there is a stringent constraint either on the accuracy of the decision to be taken, or on the energy consumption, or both. In these cases, the Edge Intelligence (EI) paradigm \cite{zhou_edge_2019} \textcolor{black}{offers a solution through computation offloading from the Edge Device (ED) to an Edge Server (ES)}, thus enabling learning and inference tasks under strict energy, latency, and reliability constraints.}
\\
In an energy saving perspective, it is important to maintain the computational complexity, of the employed learning models, as low as possible. At the same time, taking into account the envisaged exponential traffic growth in the next years \cite{strinati20216g, itu2015imt}, we also need to search for new and more sustainable communication paradigms, capable to save bandwidth and power resources, while preserving the effectiveness. To this end, \textcolor{black}{Goal-Oriented Communications (GOCs) \cite{strinati20216g} represent a hot research topic, suggesting to overcome the classical Shannon paradigm \cite{shannon1948mathematical}, which focuses on recovering all the transmitted bits, irrespective of the purpose those bits are transmitted for.} Conversely, \textcolor{black}{GOCs tunes the source encoding rule and adapts the transmission rate focusing directly on the task that motivated the exchange of information, while respecting task-specific performance constraints.} This approach makes possible to reach high compression degrees, and to save as much transmission resources as possible, any time an ED decides to offload computations to an ES.

From this viewpoint, the Information Bottleneck (IB) method\cite{tishby2000information}, inspired by rate-distortion theory arguments, is a theoretical framework that can be used to extract from data \textcolor{black}{(and parsimoniously encode)} those features that are \textcolor{black}{particularly} meaningful for a specific learning or inference task, \textcolor{black}{and, consequently, it is a promising way to formalize and implement GOCs}.
Specifically, the IB method is designed to retrieve a representation of an input random source that, for any (compact) features size \textcolor{black}{and its associated bit-compressed representation}, is as much informative as possible with respect to the target of a specific learning or inference task \cite{tishby2000information}.
Specifically, the IB acts as a supervised Feature Extractor (FE) where, differently from unsupervised methods such as PCA, the compression targets a proper representation for the final decision, rather than the fidelity of the input reconstruction. Unfortunately, since the IB relies on a functional optimization problem based on Mutual Information (MI), it admits a closed-form solution only on specific cases \cite{tishby2000information}. The most noticeable one is when the input and the output of a regression task are characterized by a joint multi-variate Gaussian distribution \cite{chechik2003information, pezone2022goal}. 
The aim of this paper is to overcome this limitation, proposing a quite general framework to adapt the Gaussian IB principle to tasks that deviate from multivariate Normal regression, \textcolor{black}{which turns out to be particularly helpful to deploy Edge Intelligence tasks using a GOC philosophy.}

\textbf{Related Works}. There are several works in the recent literature where the IB principle is employed to support ML applications \cite{zaidi2020information}. The authors in \cite{wang2023information,zhang2023multi} propose IB-based approaches for medical imaging classification and segmentation. To overcome the difficulty to adapt the IB principle without Gaussian assumptions, their solution relies on a Variational Information Bottleneck (VIB) framework \cite{alemi2016deep} that, considering a variational bound of  the IB cost function, derives a loss function that is employed to train Variational Auto-Encoders (VAEs) \cite{kingma2013auto}. 

\textcolor{black}{In \cite{Mahvari202137}, the authors propose a variation of the original Information Bottleneck problem named Scalable Information Bottleneck, where multiple compressed representations, with increasingly richer features, are considered. The work in \cite{estella2018distributed} proposes a solution for a distributed implementation of the IB problem, that is suitable in both the discrete and Gaussian case.}

Other noticeable examples can be found in \cite{alemi2018uncertainty,ahuja2021invariance}, where the IB is proposed as a method to improve the generalization capabilities for DNN-based tasks, also in the presence of out-of-distribution (OOD) data. The IB principle has been applied also in conjunction with Graph Neural Networks (GNNs), in the so-called Graph Information Bottleneck framework \cite{wu2020graph,yu2020graph}.

Some recent works testify the importance of the IB method for GOCs communications. Specifically, the approaches proposed in \cite{shao2021learning, shao2022task, chamain2022end, liu2022task,alhussein2023dynamic} can be exploited in Edge Intelligence settings, where one or more Edge Devices offload a specific learning/inference task towards the servers placed at the edge network. The work in \cite{li2023task} proposes an IB framework for task-oriented communications, which is based on a slight modification of the IB formulation to cope with OOD data. Reference \textcolor{black}{\cite{xie2023robust} proposes a Robust Information Bottleneck (RIB) formulation to cope with digital communication schemes}. In \cite{pezone2022goal}, the authors propose an optimal resource allocation framework based on the Gaussian IB, e.g., regression tasks, for Edge Machine Learning applications. \textcolor{black}{The authors in \cite{zhu2023information} propose an approach to optimize the shared codebook in Type-Based Multiple Access (TBMA) based on the Information Bottleneck.}

Except for \cite{pezone2022goal}, where the Gaussian assumption holds true, all the aforementioned works are based on the VIB, which retrieves a good approximation of the IB optimal solution, paying the cost of a considerable complexity. To reduce this complexity, \cite{binucci_adaptive_2022, binucci_dynamic_2022} propose an heuristic approximation of the IB principle, where the latent (and tunable) compact representation, together with the final decisions, are obtained by a framework that exploits multiple couples of Convolutional Encoders (CEs) and Convolutional Classifiers (CCs). 
However, this approach is highly based on empirical considerations, \textcolor{black}{without any theoretical claim on the performance and accuracy that is possible to obtain by different compression architectures, or different feature sizes. Actually, although it is practically impossible to derive a closed-form expression to link the inference accuracy of a neural network with the input feature vectors, some sub-optimal and theoretically grounded criteria can be exploited. A possible attempt in this direction can be found in \cite{wen2023task}, where the authors study a multi-user Edge-Intelligence scenario, where the Edge Devices perform feature extraction and quantization prior to the transmission of the data towards a centralized edge-server for the final inference. Specifically, the authors in \cite{wen2023task} propose to quantify performance by the so called \textit{discriminant-gain} (i.e., the symmetrized Kullback-Leibler divergence between two classes in the Euclidean feature space).
The method we propose, if applied to a Goal-Oriented and Edge-Assisted Communication Scenario (which is out of the scope of this paper) would share some similarities with \cite{wen2023task}: indeed, both the methods rely on Gaussian assumptions of the features, and their linear extraction by either PCA in \cite{wen2023task}, or Gaussian Information Bottleneck, as we propose. Specifically, the authors in  \cite{wen2023task} propose the discriminant-gain to optimally allocate resources, such as quantization bits, under energy and latency constraints. Actually, our proposal does not consider features quantization, which could be handled  either by the approach proposed in \cite{wen2023task}, 
 or simply by exploiting mean squared error (MSE) cost functions, as we will detail in Sec. III.A. Furthermore, differently from \cite{wen2023task}, our approach is naturally oriented to \emph{dynamically} optimize also the feature's vector size, by means of Sec. III.\ref{sec:multi-class}, Fig.\ref{fig:classification_with_multiple_architectures} and \cite{binucci_adaptive_2022,binucci_dynamic_2022,binucci2023multiuser}.}

\textbf{Our contribution.} In this paper we propose a novel method to adapt the Gaussian IB (GIB) principle \cite{chechik2003information} to any \textcolor{black}{Goal-Oriented} inference task that exploits a (Deep) NN to come up with a decision. In a nutshell, our solution embeds an \emph{opportunistic} regression step between a transformation of the input data and the output of the first layer of the NN trained to perform a task  (for instance, image classification). \textcolor{black}{This opportunistic regression 
is instrumental to exploit the IB method to extract the most relevant features from the data in a principled way,} giving us the flexibility to trade the learning task accuracy with the feature's vector size. 
Differently from \cite{binucci_adaptive_2022,binucci_dynamic_2022,binucci2023multiuser}, the proposed solution can be  implemented by training and exploiting a {\it single} DNN architecture, rather than a bank of DNNs, e.g., one for each specific size of the input features.
\textcolor{black}{The advantages of our formulation with respect to the Variational Information Bottleneck implementations \cite{alemi2016deep}, which is widely considered in the recent literature, are mainly related to the inference and training complexity:
\begin{itemize}
    \item First of all, in classical VIB implementations, the size of the compressed representation is constrained by the encoder architecture. Thus, like the scheme proposed in \cite{binucci_adaptive_2022, binucci_dynamic_2022,binucci2023multiuser}, VIB requests different encoder/classifier architectures for different compression levels. Conversely, the proposed approach may possibly rely on a single (classifying)  neural network, which follows a simple compression stage and is trained once and for all. 
    \item Compared to the proposed OIB approach, the VIB encoding requests an extra neural network to produce a compact representation of the data, which has to be jointly trained with the classifying network. Furthermore, the VIB training procedure needs Monte-Carlo sampling (obtained through, the reparametrization trick) to obtain an unbiased estimation of the gradient, thus further increasing the complexity with respect to the proposed OIB.
\end{itemize}}
The simulation results testify the effectiveness of the proposed approach, showing either a performance gain with respect to other compression strategies, or lower complexity, or both. \textcolor{black}{Furthermore, the simulation results also show that the proposed approach allows to minimize the entropy of the compressed representations of the input data, with respect to competitive approaches. This means, that it is theoretically possible to search for compression schemes capable to represent the source with less bits, making our proposal particularly attractive for Edge-Assisted Goal-Oriented Communications.}

\textcolor{black}{\textbf{Outline.} The rest of this paper is organized as follows. In Sec. \ref{sec:background}, we give a general background about feature compression and we briefly recall the Information Bottleneck problem, with focus on the Gaussian case. Then, in Sec. \ref{sec:approach_description}, we describe our opportunistic approach to adapt the Gaussian IB solution to general Inference Tasks, and in Sec. \ref{sec:Simulations} we present and discuss our simulation results. Finally, Sec. \ref{sec:Conclusions} draws the conclusion and sketches the possible future research directions.}

\section{Background}\label{sec:background}

\textcolor{black}{This section provides a brief summary about feature extraction and compression, and the Information Bottleneck method. For the sake of clarity, in Tab.\ref{tab:notation_table} we report the main notation and definitions used throughout the paper.}

\begin{table}[ht]
    \caption{Main Notation and Definitions}
    {\color{black}\begin{tabular}{|c|c|}
        \hline
        \textbf{Symbol} & \textbf{Definition} \\
        \hline
        $\mathcal{D}_{tr}$,  $\mathcal{D}_{test}$ & Training and test-sets\\
        \hline
        $\mathbf{x}_{n}$, $\mathbf{y}_{n}$ & Input and output for the $n$-th sample \\
        \hline
        $\tilde{\mathbf{x}}_{n}$ & Gaussian transformation of the $n$-th sample \\
        \hline
        $\tilde{\mathbf{y}}_{n}$ & Target of the $n$-th sample of the Gaussian regression task \\
        \hline
        $n_x$, $n_y$  & Size of the input and output variables \\
        \hline
        $z_n$ & Compressed representation for the $n$-th sample \\
        \hline
        $n_z$ & Size of the compressed representation \\
        \hline
        $\rho$ & Compression ratio \\
        \hline
        $\mathbf{A}_{\rho}$ & Compression matrix \\
        \hline
     $\mathbf{\Theta}_{\rho}$ & Re-expansion matrix \\
        \hline
        $\mathbf{\Sigma}_{\cdot}$ & Covariance matrix \\
        \hline
        $\mathbf{\Sigma}_{\cdot|\cdot}$ & Conditional Covariance Matrix\\
        \hline
        $H(\cdot)$ & Differential Entropy\\
        \hline
        $I(\cdot,\cdot)$ & Mutual Information\\
        \hline
    \end{tabular}}
    \label{tab:notation_table}
\end{table}

\subsection{Feature Extraction and Compression}
\textcolor{black}{Feature-Extraction (FE) aims to extract features from the raw data that are relevant for the underlying inference task. Classical machine learning models (e.g., SVM, linear regression, etc.) envisage to perform  Feature-Extraction as a separated data pre-processing stage. Conversely, when we deal with Deep Neural Networks, FE is directly embedded in the first layers, which allows us to retrieve parsimonious data representations containing all the necessary information to fulfil the final learning/inference task.
In some cases, Feature Extraction can lead to a substantial dimensionality reduction of the input data. In this case we talk about \textit{Feature Compression}, which can be highly beneficial \textcolor{black}{in all applications where it is required to share or communicate the learned representation}, as in Edge Intelligence scenarios.}

Specifically, let us consider a data-set $\mathcal{D}\{(\mathbf{x}_{n},\mathbf{y}_{n})\}_{n=1}^{N}$, where $\mathbf{x}_{n} \in \mathbb{R}^{n_{x}}$ are input variables and $\mathbf{y}_{n} \in \mathbb{R}^{n_{y}}$ are the associated labels. Feature-Compression is a pre-processing stage that reduces the size of the input data $\mathbf{x}_{n}$ from ${n}_x$ to $n_z$, by generating a new variable $\mathbf{z}_{n} \in \mathbb{R}^{n_{z}}$, whose compression-ratio is expressed by $\rho=n_x/n_z$. 
The size reduction aims to simplify the decision function of the learning task. \textcolor{black}{Furthermore, a well-designed features extraction process should allow to capture only the information that is \textit{relevant} for the specific inference task, with a possible improvement in terms of the model generalization capabilities \cite{bishop2006pattern}}.


In general, a feature extractor is modelled through a function $\mathbf{z}_{\rho}=f_{\rho}(\mathbf{x})$. When $f(\cdot)$ is linear with respect to $\mathbf{x}$, we end up with a Linear Feature Extractor (LFE)
\begin{equation}
\mathbf{z}_{\rho}=\mathbf{A}_{\rho}\mathbf{x},
\label{eq:lfe}
\end{equation}
where $\mathbf{A}_{\rho} \in \mathbb{R}^{n_{x} \times n_{z}}$ is the \textit{compression matrix}, for a specific compression ratio $\rho$. Typically, LFEs are employed \textcolor{black}{before an inference/learning task based on  non-linear models}, e.g., a (Deep) NN \cite{leiva2007maximization}, while non-linear FE are typically used jointly with a linear discriminant function.

The compression matrix $\mathbf{A}_{\rho}$ can be designed according to different (possibly optimal) criteria. \textit{Supervised} LFEs derive the projection matrix plugging the training labels $\mathbf{y}_{n}$ in the design criteria, differently from \textit{unsupervised} LFEs, like PCA \cite{abdi2010principal}, which take into account only the training input data, identifying for instance the most relevant features to reconstruct the original input $\mathbf{x}_{n}$ from the compressed features $\mathbf{z}_{\rho,n}$. 

\textcolor{black}{In the sequel we will show how the Gaussian Information Bottleneck solution naturally leads to a supervised Feature-Compression process, which can be exploited to design parsimonious computational offloading operations in Edge-Assisted inference tasks.} \textcolor{black}{More specifically, the GIB does not only identify the best features to extract, for any given feature's size, but it also weight them in order to minimize a lower bound on the bits necessary for their representation.}

\subsection{The Information Bottleneck Method}
Let's suppose to have two random variables $\mathbf{x} \in \mathbb{R}^{n_{x}}$ and $\mathbf{y} \in \mathbb{R}^{n_{y}}$. 
Our goal is to characterize the information on $\mathbf{x}$ that is relevant about $\mathbf{y}$. In a ML setting, $\mathbf{x}$ and $\mathbf{y}$ represent the input and the output of a specific inference task, respectively. 
The IB method aims to find a \emph{probabilistic} compact representation $\mathbf{z}$ of the input variable $\mathbf{x}$, while preserving a certain amount of information about the output of the inference task $\mathbf{y}$. 
This problem can be formulated in Lagrangian form \cite{boyd2004convex} as follows \cite{tishby2000information}

\begin{mini}
    {p(\mathbf{z}|\mathbf{x})}{I(\mathbf{z},\mathbf{x})-\beta I(\mathbf{z},\mathbf{y})}{}{}.
    \label{eq:ib_problem}
\end{mini}


\noindent where $I(\cdot,\cdot)$ denotes the mutual information \cite{cover1999elements} function. More specifically, we look for a statistical mapping $p(\mathbf{z}|\mathbf{x})$ that, minimizing $I(\mathbf{x},\mathbf{z})$ encourages a \emph{probabilistic} compression on $\mathbf{z}$, while controlling through $I(\mathbf{z},\mathbf{y})$ the amount of information retained by $\mathbf{z}$ about the output $\mathbf{y}$.


When the Lagrange multiplier $\beta \to \infty$ we get a more informative representation, while at the same time penalizing the compression. Otherwise, as $\beta \to 0$, we obtain a more compact representation in probabilistic sense, sacrificing the learning performance. 

In the general case, it does not exist a closed-form solution for \eqref{eq:ib_problem}. In the discrete case, we must resort to an iterative fixed-point algorithm \cite{tishby2000information,Ugur2017349}, which suffers of a considerable computational complexity and some instability, due to the estimation of the mutual information. For the continuous case, it is possible to resort to sub-optimal VIB formulations, which are quite complex to be implemented in practice \cite{shao2021learning}. A valuable exception is represented by the Gaussian case \cite{chechik2003information}, which admits an elegant closed form solution, as we recall in the following.

\subsection{Gaussian Information Bottleneck}\label{sec:GIB}

If $\mathbf{x}$ and $\mathbf{y}$ are jointly characterized by a multivariate Gaussian distribution, i.e. $(\mathbf{x},\mathbf{y}) \sim \mathcal{N}(0,\mathbf{\Sigma}_{xy})$, the solution of \eqref{eq:ib_problem} can be expressed in closed-form. More specifically, it can be proved that the optimal statistical mapping is linear, and it is given by \cite{chechik2003information}
\begin{equation}
    \label{eq:bottleneck_transform}
    \mathbf{z}=\mathbf{A}_{\rho}\mathbf{x}+\mathbf{\xi},
\end{equation}
where $\mathbf{\xi} \sim \mathcal{N}(\mathbf{0},\mathbf{I})$ is an additive Gaussian vector, statistically independent of $\mathbf{x}$ and $\mathbf{z}$, and the compression matrix $\mathbf{A}_{\rho}$ is built from the eigenvalues and left eigenvectors $\{\lambda_{i},\boldsymbol{v_{i}}\}_{i=1,\ldots,n_x}$ of $\mathbf{\Sigma}_{x|y} \mathbf{\Sigma}_{x}^{-1}$, \textcolor{black}{where $\mathbf{\Sigma}_{x|y}$ is the conditional covariance matrix of $\mathbf{x}$ given $\mathbf{y}$, while $\mathbf{\Sigma}_{x}^{-1}$ is the inverse covariance matrix of $\mathbf{x}$}. Interestingly, the solution relies on the same eigenvectors used in Canonical Correlation Analysis (CCA) \cite{hardoon2004canonical}. More specifically, sorting the eigenvectors $\boldsymbol{v_{i}}$ by the ascending values of the corresponding eigenvalues $\lambda_{i}$, for any fixed value of the Lagrange multiplier $\beta$, the matrix $\mathbf{A}_{\rho}$ is obtained by the non-zero rows of 
\begin{equation}
\label{eq:GIB-CompressionMatrix}
  {\mathbf{A}} = 
   \begin{cases}
      [\boldsymbol{0^{T}};\boldsymbol{0^{T}};\dots;\boldsymbol{0^{T}}], & 0 \leq \beta \leq \beta_{1}^{c} \\
      [\alpha_{1}\boldsymbol{v_{1}^{T}};\boldsymbol{0^{T}};\dots;\boldsymbol{0^{T}}], & \beta_{1}^{c} < \beta \leq \beta_{2}^{c} \\
      [\alpha_{1}\boldsymbol{v_{1}^{T}};\alpha_{2}\boldsymbol{v_{1}^{T}};\boldsymbol{0^{T}};\dots;\boldsymbol{0^{T}}], & \beta_{2}^{c} < \beta \leq \beta_{3}^{c} \\
        \vdotswithin{ = } &  \vdotswithin{ = } \\   
      [\alpha_{1}\boldsymbol{v_{1}^{T}};\alpha_{2}\boldsymbol{v_{2}^{T}};\dots;\alpha_{n}\boldsymbol{v_{n}^{T}}], & \beta_{n}^{c} < \beta 
    \end{cases}       
  \end{equation}

        
        

where $\beta_{i}^{c}=\frac{1}{1-\lambda_{i}}$, $\alpha_{i}=\sqrt{\frac{\beta(1-\lambda_{i})-1}{\lambda_{i}r_{i}}}$, and $r_{i}=\boldsymbol{v_{i}^{t}}\mathbf{\Sigma}_{x}\boldsymbol{v_{i}}$.

The structure of the matrix $\mathbf{A}$ makes evident the trade-off between the features' compression and the learning performance. Indeed, considering that only the first $n_z= n_x/\rho$ non-zero rows are used for compression when $\beta_{n_z}^{c} < \beta \leq \beta_{n_z+1}^{c}$, low values of $\beta$ induces the use of few CCA eigenvectors, thus reaching an extreme compression degree, with a consequent degradation in terms of learning performance. On the other hand, as $\beta$ increases, we start adding more eigenvectors to the compression matrix $\mathbf{A}_{\rho}$, thus obtaining a larger intermediate representation $\mathbf{z}_{\rho}$, with a higher amount of information with respect to $\mathbf{y}$.

\subsection{Relationships between GIB and CCA}
\label{sec:rel_gib_pca}
\textcolor{black}{
Looking at eq. \eqref{eq:GIB-CompressionMatrix} we note that, differently from CCA, the GIB solution characterizes also the loadings $\alpha_{i}$. Furthermore, it considers a noise term $\mathbf{\xi}$ in the mapping. These differences arise since the IB aims to optimize the trade-off between \textit{compression}, captured by the term $I(\mathbf{x},\mathbf{z})$, and relevance with respect to the outcome, captured by $I(\mathbf{z},\mathbf{y})$.
}
\textcolor{black}{
\textcolor{black}{Conversely, if we fix a priori the dimension $n_{z}$ of the feature vector,} it makes sense to search for the best linear mapping $\mathbf{z}=\mathbf{M} \mathbf{x} + \mathbf{\epsilon}$  that maximizes $I(\mathbf{z},\mathbf{y})$. Solving the optimization problem under Gaussian assumptions, turns out that the optimal transformation matrix $\mathbf{M^*}$ is given by the eigenvectors of $\mathbf{\Sigma}_{x|y} \mathbf{\Sigma}_{x}^{-1}$. Furthermore, $\mathbf{\Sigma_{\epsilon}}=\mathbf{0}$ (cf. \ref{sec:proof_mi} for the proof). Interestingly, the (accuracy) optimal solution does not depend on the loadings $\alpha_i$ of the eigenvectors (and is equivalent in this sense to usual CCA). This fact leads us to the following considerations:
\begin{itemize}
    \item Under Gaussian assumptions, the projection on the CCA basis vector is the best linear projection in terms of mutual information $I(\mathbf{z},\mathbf{y})$, \textcolor{black}{i.e., accuracy-wise}. 
    \item Since the solution does not depend on the loadings $\alpha_i$ on \eqref{eq:GIB-CompressionMatrix} (see \ref{sec:proof_mi}), if we focus only on the size of the compressed representation $\mathbf{z}$ without care about its representational complexity, GIB and CCA become equivalent. 
\end{itemize}
Thus, in performing feature extraction with a fixed number of components and a fixed encoding, if Multivariate Gaussian assumptions hold true, we can get rid of the noise term $\mathbf{\epsilon}$ and project on the basis vectors of CCA. \textcolor{black}{This is actually obvious, because the noise term of the GIB is not informative with respect to the task outcome $\mathbf{y}$}.  
}
 \textcolor{black}
 {Conversely, the GIB mapping in \eqref{eq:bottleneck_transform} gives us representations that are optimized in terms of minimum $I(\mathbf{x},\mathbf{z})$, \textcolor{black}{i.e., offering the possibility to design an encoding rule that minimizes the number of bits \cite{cover1999elements}, in the spirit of rate-distortion theory arguments. While this aspect may be of relative importance in classical ML, which tipically focuses only on the feature's size $n_z$}, it turns out particularly useful in Edge Inference scenarios assisted by Goal-Oriented communications, where we want to save as much transmission resources \textcolor{black}{(e.g., source-coding bits)} as possible, while targeting specific requirements in terms of learning performance \cite{strinati20216g}. 
}
\section{Proposed approach}\label{sec:approach_description}

\begin{figure*}[ht]
    \centering
    \includegraphics[width=0.75\linewidth]{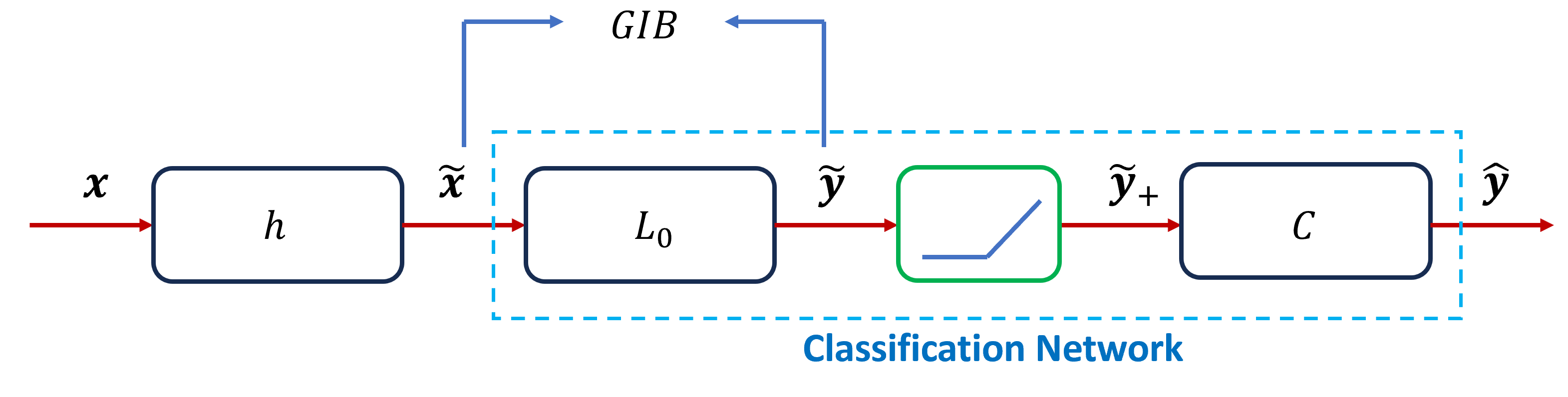}
    \caption{GIB architecture. $X$ is the input of the specific inference task (e.g., image classification), $\tilde{X}$ represents the Gaussian Transformation of the data while $\tilde{Y}$ is the output of the first layer of the considered neural network.}
    \label{fig:gib_scheme}
\end{figure*}

Herein we present a strategy to adapt the GIB framework to a generic \textcolor{black}{(non-linear, non-Gaussian)} inference task. The proposed system model is depicted in Fig.\ref{fig:gib_scheme}. We start from the input  $\mathbf{x}$ of the inference task and then we apply a linear and invertible data transformation $h(\cdot)$ to obtain another vector $\tilde{\mathbf{x}}=h(\mathbf{x})$, which is assumed to be well approximated by a multivariate Normal, by Central Limit Theorem (CLT) arguments \cite{bandyopadhyay2009asymptotic}. Without restriction of generality, in this manuscript we employ a (bidimensional) Discrete Fourier Transform (DFT) \cite{lim1990two}, which enjoys the appealing feature of fast and low complexity implementation by Fast Fourier Transform (FFT) algorithms. Thus, $\tilde{\mathbf{x}}$ represents the input for an \emph{opportunistic} regression task, that we define on purpose to exploit the GIB framework. $\tilde{\mathbf{x}}$ is also the input to the overall classifying NN shown in Fig.\ref{fig:gib_scheme}, where $\mathbf{L}_{0}$ is an operator representing the mixing stage of the first NN layer (before the activation function). 

\textcolor{black}{In this work, we stick with a linear mapping, i.e., $\mathbf{L}_{0}$ is a matrix, because it preserves the joint Gaussianity of $\tilde{\mathbf{x}}$ and $\tilde{\mathbf{y}}$: this is instrumental to exploit the GIB principle to infer $\tilde{\mathbf{y}}$ by a compressed version $\mathbf{z}_{\rho}$ of $\tilde{\mathbf{x}}$. This is consistent with the use of a shallow neural networks, as well as of a convolutional neural networks \cite{Goodfellow-et-al-2016}.}

Denoting the training-set as $\mathcal{D}_{\text{tr}}\!=\!\{(\mathbf{x}_{n},\mathbf{y}_{n})\}_{n=1}^{N_{\text{tr}}}$, the proposed training procedure is essentially divided in three steps:

\begin{enumerate}
    \item We firstly train the preferred inference network on the (transformed) training set $\{(\tilde{\mathbf{x}}_n, \mathbf{y}_n)\}_{n=1}^{N_{\text{tr}}}$\footnote{ We stress that the method can be applied to any inference/classification network. Thus, herein we do not focus on any sort of optimality of the network, and we will mostly use convolutional NN networks, due to the image classification application we focus on to verify the results.}.
    \item On the trained model, we compute the transformation $\tilde{\mathbf{y}}_n=\mathbf{L}_0\tilde{\mathbf{x}}_n + \lambda \,\mathbf{\eta}$, where $\mathbf{\eta} \sim \mathcal{N}(\mathbf{0},\mathbf{I}_{n_y})$, and $\lambda$ is a regularization parameter to be chosen. This way, we construct a regression data-set $\tilde{\mathcal{D}}_{\text{tr}}\!=\! \{(\tilde{\mathbf{x}}_n,\tilde{\mathbf{y}}_n)\}_{n=1}^{N_{\text{tr}}}$.
    \item We use $\tilde{\mathcal{D}}_{\text{tr}}$ to compute the sample-mean estimators of the CCA covariance matrices $\mathbf{\Sigma}_{\tilde{x}}$ and $\mathbf{\Sigma}_{\tilde{x}|\tilde{y}}$, that the GIB exploits to build the compression matrix $\mathbf{A}_{\rho}$, as detailed in Sec. II.\ref{sec:GIB} and \eqref{eq:GIB-CompressionMatrix}.\footnote{Actually, step (2) could be avoided because $\tilde{\mathbf{y}}=\mathbf{L}_0\tilde{\mathbf{x}}_n$ and  $\mathbf{\Sigma}_{\tilde{x}|\tilde{y}}$ can be directly computed by the knowledge (estimate) of $\mathbf{\Sigma}_{\tilde{x}}$ and $\mathbf{L}_0$ \cite{kay1993fundamentals}.}
\end{enumerate}

\noindent Specifically, the noise term we add at the output of $\mathbf{L_{0}}$ is necessary since the Gaussian Information Bottleneck fails in dealing with deterministic transformations \cite{chechik2003information}. Indeed, in case of deterministic mapping, the eigenvalues of $\mathbf{\Sigma}_{x|y} \mathbf{\Sigma}_{x}^{-1}$ would assume values in the set \textcolor{black}{$[0, 1]$} \cite{hardoon2004canonical}, thus making infeasible the computation of the loads $\alpha_i$ in \eqref{eq:GIB-CompressionMatrix}.
This way, the solution is equivalent to the well-known Ridge Canonical Correlation Analysis (RCCA) \cite{vinod1976canonical}, a regularized version of the classical CCA with a penalty on the $L_{2}$ norm of the projection vectors. 

Once we learned how to compress features by the proposed Opportunistic IB (OIB), we can exploit this knowledge to classify data on the test-set $\mathcal{D}_{\text{test}}$, according to two different schemes, as detailed in the two following sub-sections.
In particular, note that, for any possible compression ratio $\rho$, the associated matrix $\mathbf{A}_{\rho}$ is simply obtained by \eqref{eq:GIB-CompressionMatrix}, selecting the first $n_z$ eigenvectors of a fixed CCA. Thus, the compressive transformation is unique, and just the loading parameters $\alpha_i$ depends on (and has to be stored for) the specific compression ratio $\rho=n_x/n_z$, i.e., for the specific  $\beta=\beta_{n_z+1}^{c}$.\footnote{Note that, as detailed in the following, herein we can handle the classification based on a \emph{size-tunable} compressed representation $\mathbf{z}_{\rho}$, by using a \emph{single} NN, rather than a bank of encoders and classifiers, as it happens in \cite{binucci_adaptive_2022,binucci_dynamic_2022}.} 
This is clearly attractive complexity wise.

\subsection{Classification with feature expansion}
\label{sec:Feature_expansion}
The first and more appealing strategy, depicted in Fig. \ref{fig:lfe_general_scheme}, envisages to use a \emph{single} NN $C(\cdot)$ to perform the inference task, i.e., the same NN for any compression ratio $\rho$ used by the OIB-based features' compression. To this end, we don't use the first layer $\mathbf{L}_0$ the original NN has been trained with, and we use the GIB-compressed representation $\mathbf{z}_{\rho}$ to produce an estimate $\tilde{\mathbf{y}}_{\rho}^{\text{(r)}}\!=\!g_{\rho}(\mathbf{z}_{\rho})$ of the $\tilde{\mathbf{y}}\!=\!\mathbf{L}_0\tilde{\mathbf{x}}$ that would have been actually obtained by using the first NN layer $\mathbf{L}_0$. This \emph{reconstructed} $\tilde{\mathbf{y}}_{\rho}^{\text{(r)}}$ is used to feed the \emph{shortened} classification network $C(\cdot)$. For any fixed $\beta$, i.e., for any fixed compression ratio $\rho=n_{x}/n_{z}$, we assume the estimator $g_{\rho}(\mathbf{z}_{\rho})$ to be linear, as expressed by

\begin{equation}
\label{eq:linear_estimator}
    {\tilde{\mathbf{y}}_{\rho}^{\text{(r)}}}=\mathbf{\Theta}_{\rho} \mathbf{z}_{\rho}.
\end{equation}
Actually, \eqref{eq:linear_estimator} includes also the optimal Bayesian MMSE estimator, when $\Tilde{\mathbf{y}}$ and $\mathbf{z}_{\rho}$ are jointly Gaussian, as it can be safely assumed for the proposed OIB, thanks to the capability of the 2D-DFT transform to generate (almost) Gaussian distributed $\Tilde{\mathbf{x}}$. 

Thus, in this case, the Bayesian optimal reconstruction matrix $\mathbf{\Theta}_{\rho}$ is given by the classical Linear (L)-MMSE estimator that, assuming zero-mean data, is expressed by \cite{kay1993fundamentals}

\begin{equation}
    \mathbf{\Theta}_{\rho} = \mathbf{C}_{\tilde{y},z_{\rho}} \mathbf{C}_{z_{\rho}z_{\rho}}^{-1},
    \label{eq:lmmse_estimator}
\end{equation}
where $\mathbf{C}_{\tilde{y},z_{\rho}}$ is the cross-correlation between $\tilde{\mathbf{y}}$ and $\mathbf{z}_{\rho}$, while $\mathbf{C}_{z_{\rho}z_{\rho}}^{-1}$ is the inverse correlation matrix of the compressed representation. In our practical setting, considering the training-set of the \textit{opportunistic} regression sub-task, $\mathcal{\tilde{D}}_{\rho,\text{tr}}=\{(\mathbf{z}_{\rho,i},\mathbf{\tilde{y}}_{i})\}_{i=1}^{N_{tr}}$, we approximate the L-MMSE by its sample-mean counterpart, i.e., by the LS expression \cite{kay1993fundamentals}. 

\begin{equation}
\label{eq:LS_estimator}
    \widehat{\mathbf{\Theta}}_{\rho} = (\mathbf{Z}_{\rho,\text{tr}}^{T}\mathbf{Z}_{\rho,\text{tr}}^{})^{-1}\mathbf{Z}_{\rho,\text{tr}}^T \Tilde{\mathbf{Y}}_{\text{tr}}^{},
\end{equation}
where the design-matrix $\mathbf{Z}_{\rho,\text{tr}}=[\mathbf{z}_{\rho,1}^{\mathbf{T}};\dots;\mathbf{z}_{\rho,1}^{\mathbf{T}}] \! \in \!\mathbb{R}^{N_{\text{tr}} \times n_{z}}$ collects the compressed representations of the training data, while $\Tilde{\mathbf{Y}}_{\text{tr}}=[\mathbf{\tilde{y}}_{\rho,1}^{\mathbf{T}};\dots;\mathbf{\tilde{y}}_{\rho,1}^{\mathbf{T}}] \!\in \!\mathbb{R}^{N_{\text{tr}} \times n_{y}}$ collects the associated outcomes.\footnote{Note that we are implicitly assuming to cope with an over-determined system (i.e., $N_{tr}>n_{z}$) with $\mathbf{z}_{\rho,\text{tr}}$ full-rank matrix, which guarantee the uniqueness of the estimation. Otherwise, although non-unique, the L-MMSE can still be computed via Singular Value Decomposition \cite{kay1993fundamentals}. } 


It is interesting to observe that MMSE estimation of $\Tilde{\mathbf{y}}$ is also meaningful from a mutual information perspective, because for any estimator it holds true \cite{cover1999elements}
\begin{equation}
\label{eq:MSE_Inequality}
    \mathbb{E}\{||\tilde{\mathbf{y}}-\tilde{\mathbf{y}}_{\rho}^{\text{(r)}}(\mathbf{z}_{\rho})||^2\} \geq \frac{n_{y}}{2\pi e}e^{\frac{2H(\tilde{\mathbf{y}}|\mathbf{z}_{\rho})}{n_y}},
\end{equation}
where $H(\tilde{\mathbf{y}}|\mathbf{z}_{\rho})$ is the conditional entropy of $\tilde{\mathbf{y}}$ given $\mathbf{z}_{\rho}$.
Thus, considering the mutual information $I(\mathbf{z}_{\rho};\mathbf{y})=H(\mathbf{y})-H(\mathbf{y}|\mathbf{z}_{\rho})$ of the NN output $\mathbf{y}$ and the compact features $\mathbf{z}_{\rho}$, by the information processing inequality $I(\mathbf{z}_{\rho};\tilde{\mathbf{y}}) \ge I(\mathbf{z}_{\rho};\mathbf{y})$ \cite{cover1999elements}, we can conclude that the MSE minimization in \eqref{eq:MSE_Inequality}, corresponds to a maximization of an upper bound of the mutual information $I(\mathbf{z}_{\rho};\mathbf{y})$, that is highly related to the final classification accuracy, \textcolor{black}{which we actually target as the final classification perfromance.}  

This way, we pass to the classifying network $C(\cdot)$ a reconstructed $\Tilde{\mathbf{y}}_{\text{rec}}$ that is optimized from the mutual information perspective: this would possibly lead to the best possible reconstruction from the (final) classification performance point of view (see the interplay of mutual information and cross-entropy  \cite{larkin2016reflections}). \textcolor{black}{The main steps of the proposed approach are summarized in Algorithm \ref{alg:oib_algorithm}}.

\begin{algorithm}[ht]
\caption{\textcolor{black}{Opportunistic Information Bottleneck}}
\label{alg:oib_algorithm}
\begin{algorithmic}[1]
\color{black}
  \STATEnonum \textbf{Input: }\\
  \STATEnonum Training-set: $\mathcal{D}_{\text{tr}}(\mathbf{x}_{n},\mathbf{y}_{n})_{n=1}^{N_{\text{tr}}}$\\
  \STATEnonum inference network $\mathcal{M}=\mathbf{L_0} \cup [\mathbf{L_{1}},\mathbf{\dots,L_{k}}]$\\
  \STATEnonum The size of the compressed representation $n_{z}$.\\
  \STATEnonum \textbf{Output: }\\ The optimal compression transformation $\mathbf{A}_{\rho}$
  \STATEnonum \textbf{Process:}\\
  \STATE Compute a gaussian transformation $h(\cdot)$ of the input data and obtain $\{\mathbf{\tilde{x}_n}\}_{n=1}^{N_{\text{tr}}}$
  \STATE Train the preferred inference network $\mathcal{M}$ on the training-set $\{(\tilde{\mathbf{x}}_n, \mathbf{y}_n)\}_{n=1}^{N_{\text{tr}}}$\\
  \STATE Run the network to compute $\tilde{\mathbf{y}}_{n}=\mathbf{L}_0\tilde{\mathbf{x}}_{n}+\lambda \mathbf{\eta}$ and build the regression data-set $\mathcal{\tilde{D}}_{\text{tr}}=\{(\tilde{\mathbf{x}}_n,\tilde{\mathbf{y}}_{n})\}_{n=1}^{N_{\text{tr}}}$.
  \STATE Compute the Gaussian IB transformation $\mathbf{A}_{\rho}$ to infer the regression data-set $\tilde{D}_{\text{tr}}$ from a compressed representation $\mathbf{z} \in \mathbb{R}^{n_{z}}$.
  \STATE Collect the transformation for each training sample on the design matrix $\mathbf{Z}_{\rho,\text{tr}}=[\mathbf{z}_{\rho,1}^{\mathbf{T}};\dots;\mathbf{z}_{\rho,1}^{\mathbf{T}}] \! \in \!\mathbb{R}^{N_{\text{\text{tr}}} \times n_{z}}$, and the associated outcomes in $[\mathbf{\tilde{y}}_{\rho,1}^{\mathbf{T}};\dots;\mathbf{\tilde{y}}_{\rho,1}^{\mathbf{T}}] \!\in \!\mathbb{R}^{N_{\text{\text{tr}}} \times n_{y}}$.
  \STATE compute the L-MMSE estimator using Eq.\eqref{eq:lmmse_estimator} 
  \RETURN $\mathbf{A}_\rho$, $\mathbf{\tilde{\Theta}_{\rho}}$.
\end{algorithmic}
\end{algorithm}

\subsection{Classification on the compressed representation}\label{sec:multi-class}

\begin{figure*}[ht]
    \centering
    \subfloat[Features re-expanded classification with a single-NN architecture]{
        \includegraphics[width=.75\textwidth]{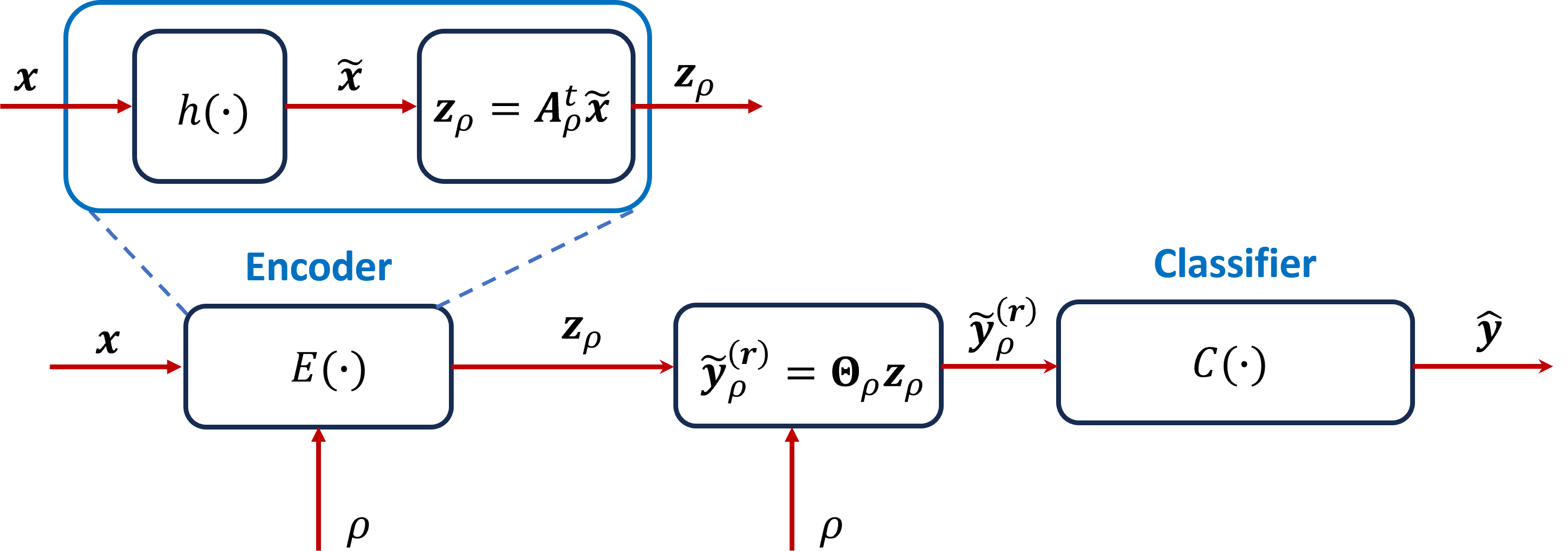}
        \label{fig:lfe_general_scheme}}

    \subfloat[Features classification with \emph{bank}-of-NN architecture.]{
        \includegraphics[width=.75\textwidth]{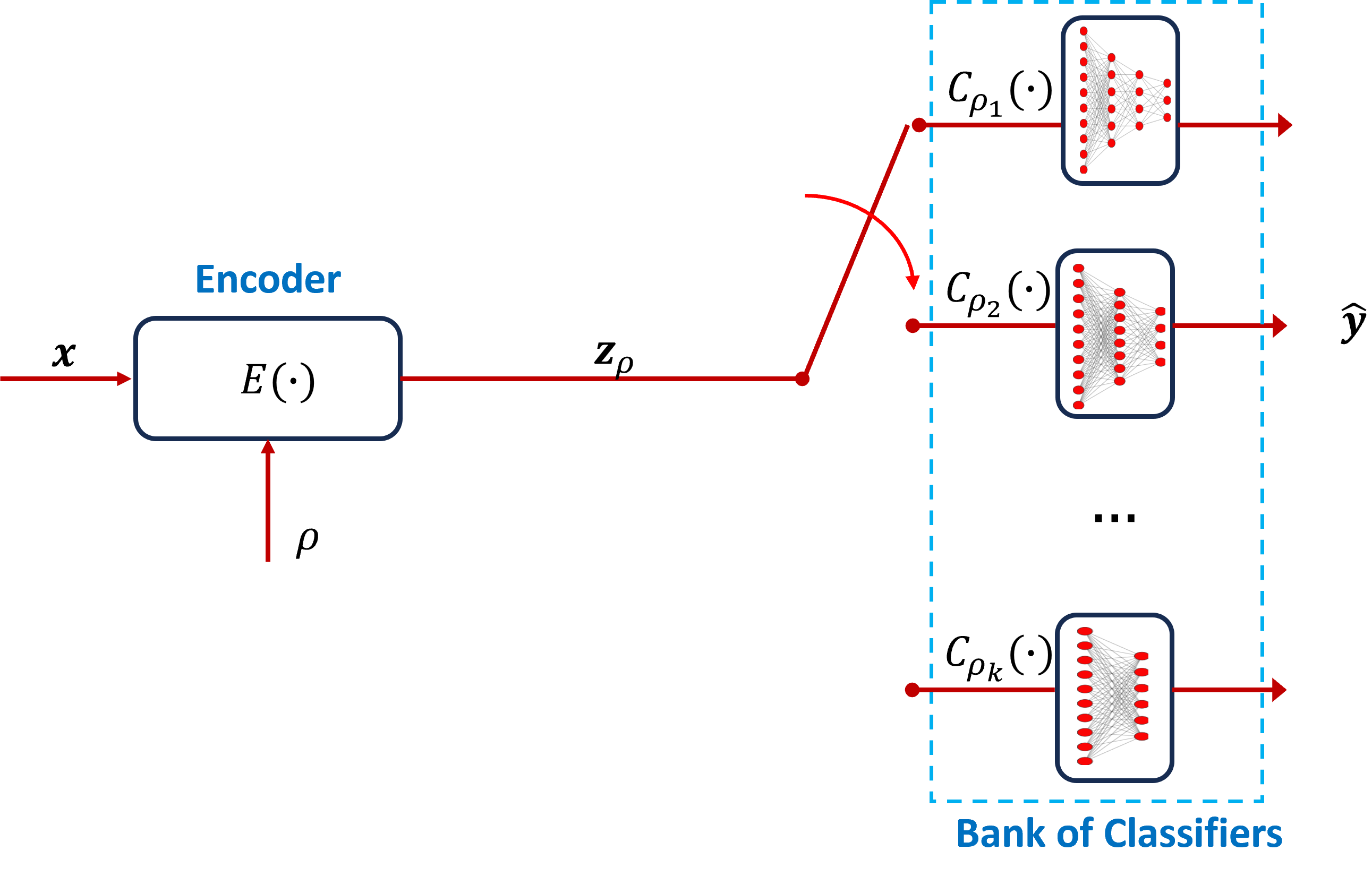}}
        \label{fig:classification_with_multiple_architectures}
    \caption{Alternative classification schemes. Note that although each LFE method shares the same linear structure of the encoder $E(\cdot)$, it is characterized by a specific (linear) transformation $h(\cdot)$ and compression matrix $\mathbf{A}_{\rho}$.}
    \label{fig:classification_schemes}
\end{figure*}

An alternative, although less flexible and appealing scheme, can be obtained assuming to use the OIB-compressed features $\mathbf{z}_{\rho}$ to directly feed a classifying NN with a (smaller) input-size $n_z$, as shown in Fig. \ref{fig:classification_with_multiple_architectures}. In this case, once we learned the OIB compression matrix $\mathbf{A}_{\rho}$ by the procedure described in the previous section, we have to train (and exploit) a specific (different) classification network $C_{\rho}(\mathbf{\cdot})$ for each possible compression factor $\rho$, ending-up with a \emph{bank} of NN classifiers, similarly to what proposed in \cite{binucci_adaptive_2022} to decode the output of a matched bank of encoders, which played the role of the OIB-encoders we introduce herein.
This further training of the bank of \emph{reduced} networks on the compressed features of the training set, may clearly help to improve the classification performance of the system, which exploits a dedicated NN for each specific compression ratio $\rho$. The price to be paid is a much higher complexity, both for the training phase, as well as for the overall architecture, which requires a bank of multiple NNs to perform the final classification. In this sense we consider this option less attractive, although it is interesting to investigate the merits and limits of the single-NN architecture we described before. 

\textcolor{black}{\subsection{Complexity Analysis}
We detail herein the complexity analysis of the proposed algorithm both for classification training and test. 
As far as the training phase is concerned,
we bear four main costs:
\begin{enumerate}
    \item The training cost of the preferred inference network, which depends on the network architecture and on the data-set.
    \item The transformation cost associated with the 2D-FFT of the training-set, which is equal to $\mathcal{O}(N_{\text{tr}}n_{x}\log_{2}{n_{x}})$.
    \item The computation cost to determine the GIB compression matrix $\mathbf{A}_{\rho}$, which is dominated by the estimation of the conditioned data covariance matrix $\Sigma_{x|y}$ and the computation of the Singular Value Decomposition (SVD) of $\mathbf{\Sigma}_{x|y}\mathbf{\Sigma}_{x}^{-1}$, leading to an overall computational complexity$\mathcal{O}(n_x^3+n_x^2N_\text{tr})$. 
    \item The computation cost of the L-MMSE estimator, which requests $\mathcal{O}(n_z^2N_\text{tr})$ operations. 
\end{enumerate}
Since the computation cost of the L-MMSE estimator and of the compression matrix $\mathbf{A}_{\rho}$ dominate the cost of the 2D-FFT, the overall computation complexity is similar to that one of a classical PCA, which also requests SVD. Furthermore, the complexity is drastically reduced with respect to the Linear Feature Extractor based on Mutual Information \cite{leiva2007maximization}, which also exploits a preliminary PCA computation, followed by a gradient based procedure to retrieve the compressed features.
During the inference procedure, all the methods based on linear feature extraction need for a computational complexity which is $\mathcal{O}(n_xn_z + n_zn_{\tilde{y}})$ due to the compression and reconstruction.On the other hand, The Variational Information Bottleneck needs an extra NN to perform compression, whose computational complexity would be dramatically higher with respect to performing a matrix multiplication. This confirms that the proposed method would be particularly useful in Edge Inference scenarios, where devices are typically characterized by limited computational capabilities.}

\section{Experiments}
\label{sec:Simulations}

In this section we discuss our experimental results to demonstrate the effectiveness of the proposed approach. Firstly, we describe the data-sets employed in our simulations. Then, for the sake of comparisons, we briefly summarize four alternative FE strategies available in the literature. Finally, we asses the performance of the proposed method.

\subsection{Data-set and NN training}

We considered three data-sets in our experiments:
\begin{itemize}
    \item The GTSRB data-set \cite{GTSRB2011}, composed of 1213 RGB traffic sign images divided in 43 different classes with a size $n_{x}=32\times32$ px $\times$ 3 colors. In order to obtain a reliable estimation of the covariance matrices involved in the OIB framework (and in PCA), we considered a data-augmentation factor equal to $5$, obtained by applying random rotations to the original images. This procedure led us to deal with an augmented data-set  of 6.065 images, that we split in 4.852 for the training-set and 1.213 for the test-set.
    \item  \textcolor{black}{The Euro-SAT data-set \cite{helber2019eurosat}, composed of \num{2.7e4} RGB images captured by the the Sentinel-2 satellite divided in 10 different classes. Also this data-set has images with size $n_{x}=32\times32$   px $\times 3$ colors. We considered 21.600 images for the training-set and 5.400 images for the test-set.}
    \item The MNIST data-set \cite{deng2012mnist}, which is composed of $\num{7e4}$ grey-scale images of handwritten digits divided in 10 different classes, with a size  $n_x=28\times28=784$ px. 
The images are split in $\num{6e4}$ for the training-set and $\num{1e4}$ for the test-set.  
\end{itemize}

We focused on these well known data-sets, because they lead to a training phase of the proposed architecture, with reasonable complexity and computational time. 
Indeed, as clarified in Sec. \ref{sec:approach_description}, the proposed OIB needs to estimate data covariance matrices (and an inverse), e.g., $\mathbf{\Sigma}_{x|y}$, and $ \mathbf{\Sigma}_{x}^{-1}$, with a complexity that scales cubical with the image size: thus, for larger images the computational burden could become problematic for a standard PC, as the one we used to test the proposed approach\footnote{The reduction of the complexity to train the system, and to estimate the covariance and precision matrices by exploiting the data-structure and parallel computations, will be investigated in future works.}.
\textcolor{black}{Regarding the training procedure, we employed the well-know \textit{Adam} optimizer \cite{kingma2014adam}. We considered 30 epochs in all our simulations, a learning rate $lr=10^{-3}$ and a batch-size $|\mathcal{B}|=32$.}

The validity of the multivariate Normal assumptions under the proposed OIB strategy, have been successfully verified  by the \textit{Henze-Zirkler} test for Multivariate Normality \cite{henze1990class}, that has been applied on the {2D-DFT} outputs $\tilde{\mathbf{x}}$ that feed the OIB stage.

\subsection{Competitive approaches}
We compare the OIB framework with four alternative FE strategies:
\begin{itemize}
    \item The well-know Principal Component Analysis (PCA) \cite{abdi2010principal}, which is typically used for \textit{unsupervised} LFE, i.e., without taking into account relevance to the specific inference task $\mathbf{y}$. In this case, for any compression ration $\rho=n_x/n_z$, the features are obtained projecting the data $\mathbf{x}$ on the first $n_z$ eigenvectors of the estimated covariance matrix $\tilde{\mathbf{\Sigma}}_{x}$.
    \item The Mutual Information Based LFE (MIB-LFE)\cite{leiva2007maximization}, which proposes to improve PCA by embedding the full PCA on a lower dimensional space by an orthogonal projection matrix $\boldsymbol{A}$, which is learnt (together with the classifying NN) by maximizing an approximated  bound of the mutual information $I(\mathbf{z}_{\rho},\mathbf{y})$. The optimization procedure is based on gradient ascent and Gram-Schmidt ortho-normalization. This method, as the proposed OIB, represents a theoretically principled, task-oriented, supervised LFE, whose effectiveness has been experimentally confirmed and validated in many works \cite{chumerin2006comparison}.
    \item The nonlinear-FE framework proposed in \cite{binucci_adaptive_2022, binucci_dynamic_2022, binucci2023multiuser}, that exploits a bank of encoders and classifiers, each characterized by a different compression ratio $\rho$. Specifically, each encoder produces $\mathbf{z}_{\rho}$ by down-sampling the input image $\mathbf{x}$ by a cascade of convolutional/max-pooling layers and it is jointly trained with an associated CC, which performs the final classification.
    \item The Variational Information Bottleneck (VIB) method \cite{alemi2016deep}, which is a well-know \emph{approximation} of the IB principle, based on a variational bound of the IB cost function in \eqref{eq:ib_problem}. Actually, also the VIB can be  implemented by considering an architecture with an encoder to extract the features $\mathbf{z}_{\rho}$ and a NN decoder to classify them \cite{alemi2016deep},\cite{shao2021learning}. The main disadvantage of the VIB is that, since the output size of the encoder is fixed, similarly to the approach described in \cite{binucci_adaptive_2022}, it is necessary to train a bank of encoders/classifiers, each one associated to a specific compression factor.
    
\end{itemize}
We stress that the first two LFE methods share a similar structure with the proposed OIB in Fig. \ref{fig:classification_schemes}. Indeed, while $h(\cdot)$ stands for 2D-DFT in the OIB, it represents a lossless PCA for the other two methods. Then, $\mathbf{A}_{\rho}=[\mathbf{I}_{n_z},\mathbf{0}]$ for unsupervised PCA, while $\mathbf{A}_{\rho}$ is a full $n_x \times n_z$ mixing matrix for MIB-LFE \cite{leiva2007maximization}. 
Consequently, like the OIB, also the first two methods admit a single-NN classification strategy, while the nonlinear-FEs in \cite{binucci_adaptive_2022, binucci_dynamic_2022, binucci2023multiuser} and the VIB are structurally designed with a bank-of-NNs, as shown in Fig.\ref{fig:classification_schemes}(b).

\subsection{Classification by a single-NN}
\label{sec:single-NN-comp}
A single-NN is trained by the procedure described in Sec. \ref{sec:approach_description}.
Thus, for any different compression ratio $\rho$, the input of this NN is obtained from the compact representation $\mathbf{z}_{\rho}$, by the LS-estimator in \eqref{eq:LS_estimator}.
Employing CLT arguments, we can assume that, also for unsupervised PCA and MIB-LFE, the compact representation $\mathbf{z}_{\rho}$ and $\tilde{\mathbf{y}}$ are jointly-Gaussian. Thus, L-MMSE expansion from $\mathbf{z}_{\rho}$ to $\Tilde{\mathbf{y}}$ turns out to be optimal, as for the OIB.
Conversely, given the non-linear structure of the compressive encoder in \cite{binucci_adaptive_2022} and of the encoders in the VIB \cite{alemi2016deep}, we cannot assume the joint Gaussianity of $\mathbf{z}_{\rho}$ and $\tilde{\mathbf{y}}$, and an L-MMSE reconstruction of the CC input would be sub-optimal.

To explore the generality and potentials of the proposed approach, we did tests by exploiting both Shallow and Convolutional Neural Networks (SNNs and CNNs). Actually a SNN has been tested only on the MNIST data-set since, differently from the GTSRB data-set, it can be reliably classified also by simple SNN models, thanks to the low complexity of the underlying classification task. 

Tabs. \ref{tab:layers_shallow_network}, \ref{tab:layers_cnn_network} \textcolor{black}{and \ref{tab:layers_eurosat_network}} report the details for baseline NNs that we employed to implement the overall classification network shown in Fig.2, without exploiting the proposed OIB approach. Tables also report the Multiplication and Accumulation Complexity (MAC), associated to each layer. All the NNs employ a $relu(\cdot)$ activation function.

\begin{table}[ht]
    \caption{Architecture of the Shallow network used to classify the MNIST data-set.}
    \centering
    \begin{tabular}{|c|c|c|c|}
        \hline
        Layer & Input Features & Output Features & Complexity [MACs] \\
        \hline
        $\mathbf{L_{0}}$ & 784 & 256 & 200704\\
        \hline
        $\mathbf{L_{1}}$ & 256 & 128 & 32768\\
        \hline
        $\mathbf{L_{2}}$ & 128 & 64 & 8192\\
        \hline
        $\mathbf{L_{3}}$ & 64 & 16 & 1024\\
        \hline
        $\mathbf{L_{4}}$ & 16 & 10 & 160\\
        \hline 
        \multicolumn{3}{|c|}{Total Complexity [MACs]} & 242848\\
        \hline
    \end{tabular}
    \label{tab:layers_shallow_network}
\end{table}

\begin{table}[ht]
    \caption{Architecture of the Convolutional Neural Network used to classify the GTSRB data-set. Each Convolutional stage considers a $2 \times 2$ stride.}
    \centering
    \begin{tabular}{|c|c|c|c|c|}
        \hline
        Layer & Type & Output Shape & Kernel & [MACs] \\
        \hline
        $\mathbf{L_{0}}$  & Strided Conv2D& $16 \times 16 \times 1$ & $9 \times 9$ & 37969\\
        \hline 
        $\mathbf{L_{1}}$  & Strided Conv2D & $8 \times 8 \times 32$ & $3 \times 3$ & 16200 \\
        \hline
        $\mathbf{L_{2}}$  & Strided Conv2D & $4 \times 4 \times 64$ & $3 \times 3$ & 225792 \\
        \hline 
        $\mathbf{L_{3}}$  & Conv2D & $4 \times 4 \times 128$ & $3 \times 3$ & 294912 \\
        \hline
        $\mathbf{L_{4}}$  & Linear & 43 & N/D & 88064 \\
        \hline 
        \multicolumn{4}{|c|}{Total Complexity [MACs]} & 662937\\
        \hline

    \end{tabular}
    \label{tab:layers_cnn_network}
\end{table}

\begin{table}[ht]
    \caption{Architecture of the Convolutional Neural Network used to classify the Euro-SAT data-set. The first stage considers a $2 \times 2$ stride, while the max-pooling stages consider a $2\times2$ window.}
    \centering
    {\color{black}\begin{tabular}{|c|c|c|c|c|}
        \hline
        Layer & Type & Output Shape & Kernel & [MACs] \\
        \hline
        $\mathbf{L_{0}}$  & Strided Conv2D & $32 \times 32 \times 1$ & $3 \times 3$ & 24300\\
        \hline 
        $\mathbf{L_{1}}$  & Conv2D+MaxPool & $16 \times 16 \times 64$ & $3 \times 3$ & 518400 \\
        \hline
        $\mathbf{L_{2}}$  & Conv2D+MaxPool& $4 \times 4 \times 128$ & $3 \times 3$ & 14450688 \\
        \hline 
        $\mathbf{L_{3}}$  & Conv2D+MaxPool & $2 \times 2 \times 256$ & $3 \times 3$ & 10616832 \\
        \hline
        $\mathbf{L_{4}}$  & Linear & 256 & N/D & 1048576 \\
        \hline 
        $\mathbf{L_{5}}$  & Linear & 10 & N/D & 2560 \\
        \hline 
        \multicolumn{4}{|c|}{Total Complexity [MACs]} & 26661356\\
        \hline
    \end{tabular}}
    \label{tab:layers_eurosat_network}
\end{table}

\begin{table}[ht]
    \caption{Saving in terms of computational complexity for both the UE and the ES for the shallow NN architecture (MNIST data-set).}
    \centering
    \begin{tabular}{|c|c|c|c|}
    \hline
         $n_{z}$ &  Comp. [MACs] & Class. [MACs] & Comp. Save $[\%]$ \\
         \hline
         10 & 18080 & 44704 & 74.13\\
          \hline
         20 & 25920 & 47264 & 69.84\\
         \hline
         30 & 33760 & 49824 & 65.56\\
         \hline
         40 & 41600 & 52384 & 61.27\\
         \hline
         50 & 49440 & 54944& 56.99\\
         \hline
         60 & 57280 & 57504 & 52.70\\
         \hline
         70 & 65120 & 60064 & 48.42\\
         \hline
         80 & 72960 & 62624 & 44.13\\
         \hline
         90 & 80800 & 65184 & 39.85\\
         \hline
         100 & 88640 & 67744 & 35.56\\
         \hline
    \end{tabular}
    \label{tab:saving}
\end{table}

We remind that the goal of the proposed OIB is to implement classification by a \emph{size-tunable} feature extraction, optimally trading performance for complexity. The \emph{single-network} architectures summarized in Fig. \ref{fig:classification_schemes}, fulfil this goal by introducing a 2D-DFT processing, and replacing layer $\mathbf{L}_0$ by the compressing matrix $\mathbf{A}_{\rho}$ and the expanding matrix $\mathbf{\Theta}_{\rho}$. Tab. \ref{tab:complexity table} reports the MACs of the proposed architecture, for different compression ratios $\rho=n_x/n_z$, distinguishing between the compressive FE and the final classification, which exploits only layers $\mathbf{L}_1$-$\mathbf{L}_4$ of the original network. 

The FE cost is due to the 2D-FFT and the subsequent multiplication by $\mathbf{A}_{\rho}$, while the \emph{classification cost} is due to multiplication with the matrix $\mathbf{\Theta}_{\rho}$, and the computations through the NN layers $\mathbf{L}_1$-$\mathbf{L}_4$.\footnote{The proposed architecture is particularly valuable when the FE is performed at a given device, while the final classification at a central server equipped with more powerful processing capabilities, such as it happens in cloud-based applications, and Edge Machine Learning \cite{merluzzi2021wireless,pezone2022goal,binucci2023multiuser} frameworks.}

\begin{figure}[ht]
    \centering
\includegraphics[width=1.00\linewidth]{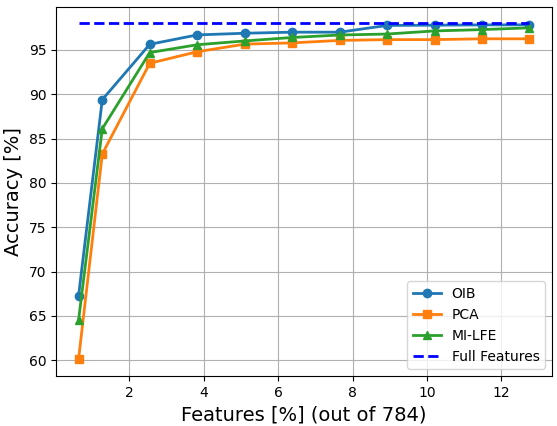}
    \caption{Performance comparison with feature's re-expansion and a \emph{single}-NN architecture.}
    \label{fig:ideal_condition_comparison}
\end{figure}

Fig.\ref{fig:ideal_condition_comparison} plots the accuracy performance of the single-NN, as a function of the number $n_z$ of extracted features, i.e., of the compression ratio $\rho=784/n_z$. It is clear how the proposed approach outperforms both the MIB-LFE \cite{leiva2007maximization} and the unsupervised PCA compression \cite{abdi2010principal}. PCA, as expected, shows the worst performance because it retrieves features that are the most informative for image \textit{reconstruction} purposes, without taking into account their relevance for the actual classification task. However, even though \cite{leiva2007maximization} would allow to extract features in a supervised and task-oriented manner, its adaptation to such a single-NN architecture, with tunable compression ratio, produces higher performance degradation with respect to the proposed OIB. 

\subsection{Analysis of compression/relevance trade-offs}
\textcolor{black}{
In the previous section, we testified the effectiveness of our method as a supervised linear feature extractor. However, as already pointed out in Sec. \ref{sec:background}.\ref{sec:rel_gib_pca}, the Information Bottleneck method, that the OIB takes inspiration from, is focused on a compression/relevance trade-off. The compression term is associated to the mutual information $I(\tilde{\mathbf{x}},\mathbf{z})$ (cf eq. \eqref{eq:ib_problem}), that the IB aims to minimize in order to retrieve a compact representation of the input, although still \textit{relevant} for the outcome of the learning task. 
Taking in mind the relationship $I(\tilde{\mathbf{x}},\mathbf{z})=H(\mathbf{z})-H(\mathbf{z}|\mathbf{\tilde{x}})$ we note that minimizing the IB cost is \textcolor{black}{somehow related to reducing the entropy of the compressed representation $H(\mathbf{z})$}, which represents a lower-bound on the minimum number of bits required to encode the source without loss of information \cite{cover1999elements}. 
Thus, in Edge-Assisted Goal-Oriented Communications scenarios, where we aim to transmit the minimum amount of data to pursue a \textit{Goal}, minimizing latency and energy consumption \cite{strinati20216g}, it makes sense to analyze the task performance (e.g., the correct classification rate) as a function of the entropy of the compressed representation.}

\begin{figure}[ht]
    \centering
    \includegraphics[width=1.00\linewidth]{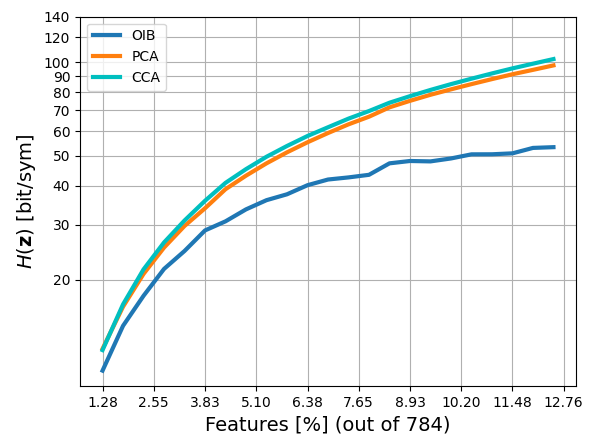}
    \caption{$H(\mathbf{z})$ as a function of the number of extracted components for PCA, CCA and OIB}
    \label{fig:entropy_vs_components}
\end{figure}

\textcolor{black}{
Furthermore, although the loading coefficients $\alpha_i$ in \eqref{eq:GIB-CompressionMatrix} of the GIB solution don't carry any improvement in terms of learning performance with respect to any other loading set, as clarified in Sec. \ref{sec:approach_description}.\ref{sec:rel_gib_pca}, they turn out useful in a data compression perspective. To better highlight this fact, we compare our OIB approach with PCA and CCA. Specifically, CCA projects the inputs $\mathbf{\tilde{x}}$ in the same directions of the OIB solution, giving however the same load to each component (i.e., $\{\alpha_{i}\}_{i=1}^{n_{z}}=1)$. 
}
\textcolor{black}{
We did our comparisons on the MNIST data-set, trained on the shallow network architecture considered in the previous section. The network considered for PCA has been trained on the original images of the data-set, while for CCA and OIB we used the same network architecture trained on the 2D-DFT of the images. 
}
\textcolor{black}{
In Fig.\ref{fig:entropy_vs_components} we show the entropy $H(\mathbf{z})$ as a function of the number of components for the considered approaches. As expected, OIB leads to a lower entropy on the compressed representation with respect to the competitors. This means that it is possible to find an encoding rule that allows to minimize the number of bits to represent $\mathbf{z}$, and associated savings in terms of transmission resources.
}

\begin{figure}[ht]
    \centering
    \includegraphics[width=1.00\linewidth]{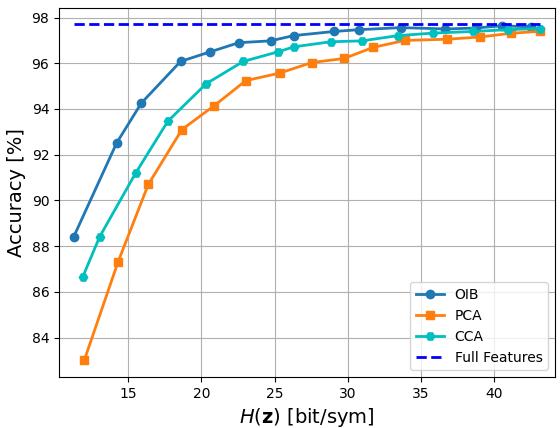}
    \caption{Accuracy as a function of $H(\mathbf{z})$  for PCA, CCA and OIB.}
    \label{fig:accuracy_vs_entropy}
\end{figure}

\textcolor{black}{
In Fig.\ref{fig:accuracy_vs_entropy} we compare the three compression philosophies considering the informativeness of the compressed representation with respect to the classification task with the same entropy amount $H(\mathbf{z})$. Using Central Limit Theorem arguments, we can assume that the compressed representation $\mathbf{z}$ has a multivariate Normal distribution. Thus, we computed the entropy considering the formula $H(\mathbf{z})=\frac{1}{2}\log((2\pi e)^{n_{z}}|\Sigma_{\mathbf{z}}|)$\cite{cover1999elements}, where the covariance matrix has been estimated by sample-mean\footnote{Note, since the differential entropy is scale-dependant, to make fair comparisons we normalized the power of $\mathbf{z}$ for all the considered approaches.}.
As expected, our OIB approach performs better with respect to the competitive approaches, since it is the only one specifically focused on the compression maximization under learning performance constraints. 
As in the previous case, the unsupervised PCA approach shows the worst performance. Regarding CCA, taking in mind that the loadings do not affect the task performance, if we consider the same compressed size $n_{z}$ we would have the same performance of the OIB approach. However, as shown in Fig.\ref{fig:entropy_vs_components}, CCA leads to a solution characterized by a higher complexity of representation (captured by $H(\mathbf{z})$) for a fixed number of components. To reach the same complexity with the OIB transformation, we have to consider more components, with a consequent improvement in terms of accuracy, when comparing the schemes at the same entropy.  
}

\subsection{Classification with a bank-of-NNs}
In this section we test the performance using the architecture in Fig. \ref{fig:classification_with_multiple_architectures}, which exploits multiple classifying NNs, each one with an input size that is matched to the size $n_z$ of the extracted features $\mathbf{z}_{\rho}$, without re-expanding them to $\tilde{\mathbf{y}}_{\rho}^{(r)}$. For the CE compression strategy, we employed the joint CE/CC training procedure described in \cite{binucci_adaptive_2022}. 
To perform fair comparisons, for the CE compression strategy we employed a bank of CEs, whose MACs complexity is similar to the OIB method. This has been done considering a mixed architecture composed of two convolutional layers followed by a fully connected layer, used to adjust the output size to the desired size $n_{z}$. 
The number of channel at the output of the first layer has been adjusted in order to match the desired complexity. 
The architecture of the encoder for the sizes $n_{z}$ is reported in Tab.\ref{tab:complexity table}. In the table we also report the comparison between the computational complexity of the proposed encoder and the Opportunistic Information Bottleneck. 
 
We assessed our performance considering the GTSRB data-set\cite{GTSRB2011} and we compare the proposed approach with the Method proposed in \cite{leiva2007maximization} and with the well-know Variational Information Bottleneck (VIB) \cite{alemi2016deep}. Indeed, these two methods based on Information Theory arguments \cite{cover1999elements}, and they are specifically tailored to classify directly on the compressed features without require the re-expansion.

\begin{table*}[ht]
    \caption{Variable encoder architecture for the different compression factors. $C_{1,2}$ represents the two convolutional layers, while $\mathbf{L_{1}}$ is the linear layer. For the convolutional layers we considered $3\!\times\!3$ convolutions and an input image with size $32\!\times\!32\times\!3$ px.}
    \centering
    \begin{tabular}{|c|c|c|c|c|c|}
    \hline
         $n_{z}$ & Output shape $C_{1}$ & Output shape $C_{2}$ & Output shape $\mathbf{L_{1}}$ & OIB [MACs] & CE [MACs]\\
         \hline
         30 & $15\!\times\!15\!\times\!15$ & $7\!\times\!7\times\!1$ & $30$ & $101408$ & $105386$\\
         \hline
         60 & $15\!\times\!15\!\times\!28$ & $7\!\times\!7\times\!1$ & $60$ & $193568$ & $196917$\\
         \hline
         90 & $15\!\times\!15\!\times\!41$ & $7\!\times\!7\times\!1$ & $90$ & $285728$ & $288448$\\
         \hline
         120 & $15\!\times\!15\!\times\!54$ & $7\!\times\!7\times\!1$ & $120$ & $377888$ & $379979$\\
         \hline
         150 & $15\!\times\!15\!\times\!67$ & $7\!\times\!7\times\!1$ & $150$ & $470048$ & $471509$\\
         \hline
         180 & $15\!\times\!15\!\times\!80$ & $7\!\times\!7\times\!1$ & $180$ & $562208$ & $563040$\\
         \hline
         210 & $15\!\times\!15\!\times\!93$ & $7\!\times\!7\times\!1$ & $210$ & $654368$ & $654571$\\
         \hline
         240 & $15\!\times\!15\!\times\!106$ & $7\!\times\!7\times\!1$ & $240$ & $746528$ & $746102$\\
         \hline
    \end{tabular}
    \label{tab:complexity table}
\end{table*}

\begin{figure}
    \centering
    \includegraphics[width=1.00\linewidth]{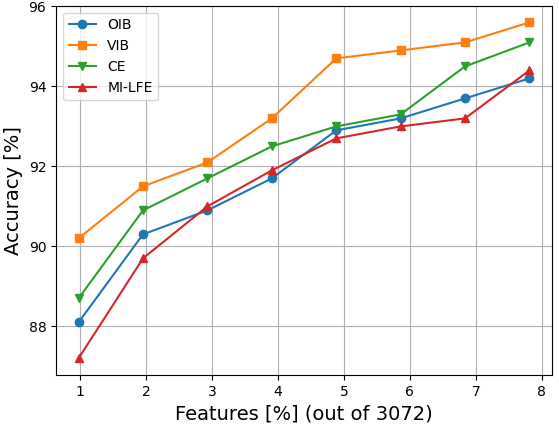}
    \caption{Performance comparison with the \emph{bank}-of-NN architecture.}
    \label{fig:performaces_different_architectures}
\end{figure}

Fig.\ref{fig:performaces_different_architectures}, which shows the accuracy versus the percentage of compressed features, witnesses that the Variational Information Bottleneck, performs slightly better with respect to the other methods. VIB represents an alternative way to train a Convolutional Encoder, which is based on a more principled approximation of the Information Bottleneck principle \cite{alemi2016deep}. Furthermore, differently from what we will show in Sec. \ref{sec:Simulations}.\ref{sec:vib_comp}, the VIB performs better with respect to the OIB framework. This represents a benefit of the joint training procedure of the compression and classification networks based on the Variational Bound \cite{alemi2016deep}, which represents a theoretically-grounded approximation of the IB objective function. Conversely, there are no substantial differences between the proposed approach, the CE-based compression and the Linear Feature Exctractor based on mutual information. As far as the CE-based compression is concerned, the CEs in \cite{binucci_adaptive_2022}, can be interpreted as a (non-linear) heuristic mimicking of the Information Bottleneck principle: somehow this fact is confirmed by the similar performance with respect to the proposed OIB, which is also a sub-optimal approximation of the IB, although more theoretically principled.
The proposed OIB in such a multi-NN implementation, performs similarly to the MIB-LFE in \cite{leiva2007maximization}, for any compression ratio. Note, that in this case we are not re-expanding the feature size. i.e., we are working for each compression ratio with the structure underlying the original MIB-LFE design. Thus, in this case OIB and MIB-LFE performs similarly because they share a quite similar structure and both the methods propose a way to approximate the maximization of the mutual information between the compact representation $\mathbf{z}_{\rho}$ and the learning task output $\mathbf{y}$. Specifically, \cite{leiva2007maximization} proposes an approximation of the $I(\mathbf{z}_{\rho},\mathbf{x})$ that  analytically relies on a statistical independence assumptions of the data, which is only approximated by the full-PCA pre-processing for non-Gaussian features. Further approximations are introduced in \cite{leiva2007maximization} to compute the \textit{negentropy} in  the objective function.
In the proposed OIB, we opportunistically embedded in the original inference task, another IB sub-problem, that we are able to optimally solve in closed form by the GIB. The main assumption here is that the good performance of the surrogate linear regression task solved by the GIB, induces also an almost optimal performance of the associated classification task. This intuitive explanation, which has been partially supported by information theoretic arguments in section \ref{sec:Feature_expansion}, is confirmed by the comparable performance with the MIB-LFE approach.
The nice part of the OIB is that it is a bit easier to be trained and furthermore, shows better performance in the single-NN architecture described in the previous section.

\subsection{Comparisons with the Variational Information Bottleneck} \label{sec:vib_comp}
To further prove the effectiveness of the proposed approach, we compare it with the well-known Variational Information Bottleneck (VIB) method \cite{alemi2016deep}. \textcolor{black}{We made our experimental validation considering the network architectures reported in Tab.\ref{tab:layers_cnn_network} and Tab.\ref{tab:layers_eurosat_network}, trained on the GTSRB\cite{GTSRB2011} and the Euro-SAT data-sets\cite{helber2019eurosat}}. VIB is a well-know approximation of the Information Bottleneck principle, based on a variational bound of the IB cost function, which can be easily implemented considering an encoder/decoder architecture.

\textcolor{black}{As already pointed out,} the main disadvantage of the VIB is that, since the output size of the encoder is fixed, similarly to the approach described in \cite{binucci_adaptive_2022}, it is necessary to train a bank of encoders/classifiers, each one associated to a specific compression factor $\rho$. \textcolor{black}{Thus, to fairly compare the two approaches, we adapted the VIB formulation to work with the single-architecture setting, considering the following training procedure:}.

\begin{enumerate}
    \item We trained the CNN reported in Tab.\ref{tab:layers_cnn_network} \textcolor{black}{{on the GTSRB data-set} the CNN reported in Tab. \ref{tab:layers_eurosat_network} on the Euro-SAT data-set}.
    \item We trained the VIB networks considering the procedure described in \cite{alemi2016deep} for all the possible number of features. We implemented an encoder with the following structure: Conv2D(F) $\rightarrow$ Linear(256$\times$F,$2 \times n_{z}$), where F is the number of output filters. The number of filters has been changed for the different compression factors to maintain the same complexity, in terms of MACs, of the OIB compression. As far as the decoder is concerned, we implemented the same architecture reported in Tabs.\ref{tab:layers_cnn_network} and \ref{tab:layers_eurosat_network}, ignoring the layer $\mathbf{L}_0$. We considered a linear layer between the encoder and the decoder in order to expand the encoder to the expected input size of the layer $\mathbf{L}_1$, i.e., $16 \times 16 \times 1$ for the network reported in Tab.\ref{tab:layers_cnn_network} \textcolor{black}{and $32 \times 32 \times 1$ for the network reported in Tab.\ref{tab:layers_eurosat_network}}. 
    \item After the end of the training procedure, we got rid of the decoder and we performed the same procedure reported in Sec.\ref{sec:approach_description} in order to get the LS estimators $\Theta_{\rho}$, which allows us to reconstruct the input of the intermediate layer $\mathbf{L}_1$ from the compressed representation obtained through the VIB encoder.
    \item We re-trained the classification networks with re-expansion for each possible compression factor $\rho$. 
\end{enumerate}

In order to fairly compare the proposed OIB with the VIB, we also re-trained the OIB classification network for each possible size of $n_{z}$ (i.e., for each $\rho$), by re-expanding the compressed features $\mathbf{z}_{\rho}$ with the proper LS estimator $\tilde{\mathbf{\Theta}}_{\rho}$. 

\begin{figure}[ht]
    \centering
    \includegraphics[width=1.00\linewidth]{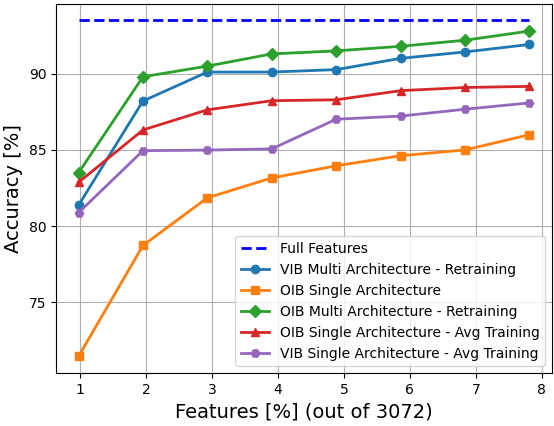}
    \caption{Comparisons between the proposed approach and the Variational Information Bottleneck \textcolor{black}{on the GTSRB data-set.}}
    \label{fig:gtsrb_comparisons}
\end{figure}

\begin{figure}[ht]
    \centering
    \includegraphics[width=1.00\linewidth]{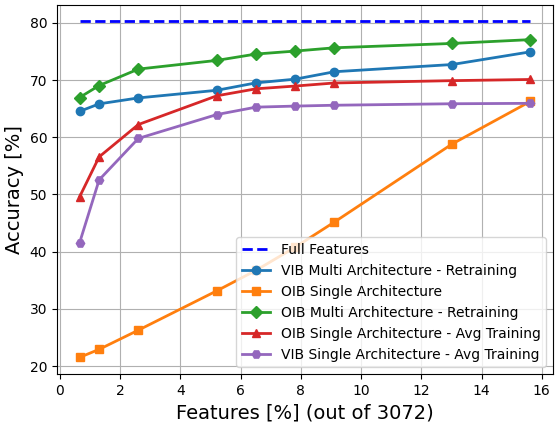}
    \caption{Comparisons between the proposed approach and the Variational Information Bottleneck \textcolor{black}{on the Euro-SAT data-set.}}
    \label{fig:eurosat_comparisons}
\end{figure}

In Fig. \ref{fig:gtsrb_comparisons} we report the average accuracy versus the percentage of compressed features on the test-set of the GTSRB data-set. We firstly note that, as expected, the re-training procedure on the expanded features allows to highly improve the performance for the proposed OIB, which slightly outperforms the classification accuracy of the VIB. However, a classical VIB approach, with a tunable FE size,  naturally leads to a higher complexity of the system, since it is necessary to consider a classification architecture with multiple NNs, each one trained on a compressed ${\mathbf{z}}_{\rho}$ with a specific number of features $n_{z}$ (i.e., a specific $\rho=n_x/n_z$). 
To avoid such implementation complexity, another possible approach is to consider a single classifying NN that is trained on a set of $\tilde{\mathbf{y}}_{\rho}^{(r)}$, which are re-expanded versions of ${\mathbf{z}}_{\rho}$, having different compression ratios, i.e., $\rho \in 
\{\rho_1,\rho_2,.....,\rho_k\}$ and, consequently, a different number of features $n_z$. Specifically, during the training phase, we randomly compressed the training data considering a uniform discrete distribution for the possible values of $\rho$. Then, we tested this \emph{average} architecture, both for the VIB and the OIB design, at different (fixed) levels of compression. Fig. \ref{fig:gtsrb_comparisons} shows that such \emph{average} classifying NN suffers some performance degradation with respect to the classification based on the multiple-NNs architectures. However, it also grants a noticeable performance improvement with respect to the single-NN architecture without re-training and, noticeably, also in this case the quite simple and elegant OIB framework outperforms the VIB.
\textcolor{black}{The same considerations hold also for the Euro-SAT data-set, as witnessed by Fig. \ref{fig:eurosat_comparisons}.}

\subsection{Sensitivity Analysis}
\textcolor{black}{This section analyzes the sensitivity of the proposed method to the neural network structure, focusing on convolutional and fully connected neural networks.
Let us consider the case of data classification with feature re-expansion (through the L-MMSE estimator).
\\
We firstly remark that we are approximating the first linear layer of our network $\mathbf{L_0}$, by a (low-rank) matrix factorization expressed by $\widetilde{\mathbf{L}}_{0}=\mathbf{A}_{\rho}\mathbf{\Theta_{\rho}}$. 
Thus, the robustness of our algorithm is related to the error we induce by said approximation, which naturally gets worse using smaller sizes $n_z$ (i.e, lower-rank approximation) for the compressed representation.
\\
However, it is important to note that the possible classification degradation depends also on another factor, i.e., the capability of the last network layers $[\mathbf{L_1},\ldots,\mathbf{L_n}]$ to be robust, or capable to adapt, to additive noise ignited in the first network layer $\mathbf{L}_0$.}


\textcolor{black}{
\indent To clarify this point, we show in the following the simulation results obtained exploiting both fully connected and convolutional neural networks, \textcolor{black}{evaluating the accuracy degradation with respect to the originally trained (full) NN, for different number $n_z$ of extracted features.} Specifically, we used the the MNIST data-set, and the system has been retrained with the procedure detailed in Sec. \ref{sec:approach_description}.A. Results are shown for a set of possible feature's size $n_z$ $\in$ $[5,10,15,\cdots,70]$.
\\
\indent 
For the fully connected neural network, the first layer is characterized by a matrix $\mathbf{L_{0}} \in \mathbb{R}^{784\times196}$, while for the convolutional architecture we employed a $2\times2$ strided convolution to down-sample the input data from $28\times28=784$ px to $196$ px, ending up with an equivalent (block diagonal) matrix $\mathbf{L_{0}} \in \mathbb{R}^{784 \times 196}$.}

\begin{figure}
    \centering
    \includegraphics[width=1.00\linewidth]{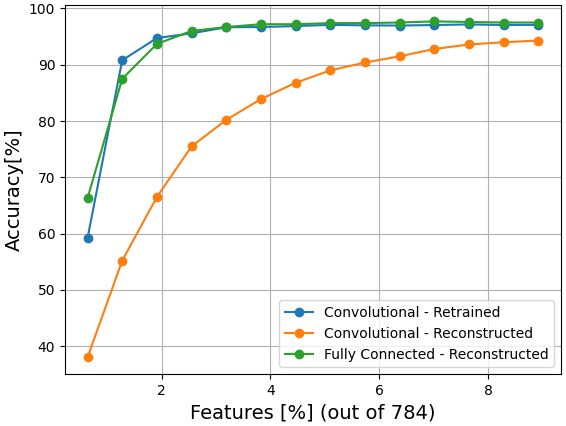}
    \caption{Accuracy on the test-set of the MNIST data-set for different number of components $n_{z}$ on shallow and convolutional neural networks.}
    \label{fig:reconstruction_error_comparison}
\end{figure}

\textcolor{black}
{As expected, Fig. \ref{fig:reconstruction_error_comparison} shows that the OIB-approximated NN provides better classification accuracy as we increase the size $n_z$ of the compressed representation $\mathbf{z}_{\rho}$, for any fixed architecture and (re-) training strategy, due to the fact that the low-rank approximation 
$\tilde{\mathbf{L}}_0=\mathbf{A}_{\rho}\mathbf{\Theta}_{\rho}$ gets better.
Furthermore, Fig.\ref{fig:reconstruction_error_comparison} shows that the convolutional neural network seems to be more sensitive to the approximation of $\tilde{\mathbf{L}}_0$. This higher sensitivity of CNN can be explained by considering that reconstructing $\tilde{\mathbf{y}}$ by $\mathbf{\tilde{L}_0}\tilde{\mathbf{x}}$, has an effect similar to adding noise on the inputs of the next network layer $\mathbf{{L}_1}$, and is well known that CNNs have robustness issues in this case \cite{hendrycks2018benchmarking}. However, this \emph{in-layer} noise effect can be mitigated by re-training the last network layers $[\mathbf{L_1},\ldots,\mathbf{L_n}]$ on the reconstructed data, i.e., on the output of $\mathbf{\widetilde{L}_0}$, paying the price to introduce some extra training complexity. 
Actually, by this retraining, the CNN performance for the proposed classification data-set are equivalent to those of the fully-connected architecture, as witnessed the blue curve in Fig.\ref{fig:reconstruction_error_comparison}.
Although, this is just a simple example, it is reasonable that properly retraining also other NN architectures, it is possible to make the approximation error introduced by OIB quite insensitive to the specific NN, as the NN layers $[\mathbf{L_1},\ldots,\mathbf{L_n}]$ should be capable to capture by re-training the mutual information $I(\tilde{\mathbf{y}}^{(r)}_{\rho},\hat{\mathbf{y}})$ if they were capable to capture $I(\tilde{\mathbf{y}},\hat{\mathbf{y}})$}.

\section{Conclusion and future work}
\label{sec:Conclusions}
We presented a new approach to opportunistically exploit the closed-form solution of the Gaussian Information Bottleneck (GIB) in a general inference task, to enable a tunable and effective supervised feature extraction, \textcolor{black}{highlighting the specific merit of the GIB with respect to CCA, as the best compact representation of the (same) extracted features}.
Experimental results on an image classification task testifies the effectiveness of the proposed approach and encourages to further investigate this research line. In particular, the proposed formulation seems particularly attractive any time there is the need, or the opportunity, to perform inference by a single-NN, with a number of features that may change in time as a function of the system resources, such as hardware, energy, storage, communications, and computations. 
In this view, future works may include the employment of the proposed compression strategy in dynamic resource allocation strategies for edge-assisted goal-oriented communications scenarios (see, e.g., \cite{binucci_adaptive_2022,binucci_dynamic_2022,binucci2023multiuser}). Furthermore, it may be interesting to apply the OIB formulation to different inference tasks, to better assess its generality. Finally, the investigation of possible interplays between OIB and Variational IB formulations represents another appealing research line. 

\section*{APPENDIX}
\label{sec:proof_mi}
\subsection*{Maximization of Mutual Information under Gaussian assumptions}

\textcolor{black}{
Let us define the following optimization problem
\begin{align}
\!\min_{\Phi_{s}}       &\quad I(\mathbf{z},\mathbf{y}) \notag
\\[0.5pt]
\text{s.t.}
&\quad \mathbf{z}=\mathbf{M}\mathbf{x}+\mathbf{\epsilon} \\
\notag
\label{eq:server_prob2}
\end{align}
}
\textcolor{black}{
where $\mathbf{x}$ and $\mathbf{y}$ are characterized by a multivariate Normal distribution \textcolor{black}{and $\epsilon$ is also Normal and statistically independent from both $\mathbf{x}$ and $\mathbf{y}$}. Since $I(\mathbf{z},\mathbf{y})=H(\mathbf{z})-H(\mathbf{z}|\mathbf{y})$, recalling the expression of the entropy for Multivariate Gaussian variables \cite{cover1999elements}, the objective function becomes
\begin{equation}
    \begin{aligned}
I(\mathbf{z},\mathbf{y})=\log(|\mathbf{M}\mathbf{\Sigma_x}\mathbf{M^t}+\mathbf{\Sigma_{\epsilon}}|)-\log(|\mathbf{M}\mathbf{\Sigma_{x|y}}\mathbf{M^t}+\mathbf{\Sigma_{\epsilon}}|).\notag
    \end{aligned}
\end{equation}
}
\textcolor{black}{
Proceeding as in \cite{chechik2003information}, nulling the derivative $\frac{\partial I(\cdot,\cdot)}{\partial \mathbf{M}}$ we obtain the following equation
\begin{equation}
\label{eq:nulling_derivative}
    \begin{aligned}
&[\mathbf{M}\mathbf{\Sigma_{x}}\mathbf{M}^t+\mathbf{\Sigma_{\epsilon}}]^{-1}\textcolor{black}{\mathbf{M}}\mathbf{\Sigma_{x}}-[\mathbf{M}\mathbf{\Sigma_{x|y}}\mathbf{M}^t+\mathbf{\Sigma_{\epsilon}}]^{-1}\textcolor{black}{\mathbf{M}}\mathbf{\Sigma_{x|y}}=0\\
    &[\mathbf{M}\mathbf{\Sigma_{x|y}}\mathbf{M}^t+\mathbf{\Sigma_{\epsilon}}][\mathbf{M}\mathbf{\Sigma_{x}}\mathbf{M}^t+\mathbf{\Sigma_{\epsilon}}]^{-1}\mathbf{M}=\mathbf{M}\mathbf{\Sigma_{x|y}}\mathbf{\Sigma_{x}^{-1}}.
    \end{aligned}
\end{equation}
}
\textcolor{black}{
Thus, $\mathbf{M}\mathbf{\Sigma_{x|y}}\mathbf{\Sigma_{x}^{-1}}$ must reside in the space generated by the rows of $\mathbf{M}$, and it is composed by left eigenvectors of $\mathbf{\Sigma_{x|y}}\mathbf{\Sigma_{x}^{-1}}$. This means we can write the optimal transformation matrix as $\mathbf{M}=\mathbf{W}\mathbf{V}$, where the rows of $\mathbf{V}$ are composed by the left eigenvector of $\mathbf{\Sigma_{x|y}}\mathbf{\Sigma_{x}^{-1}}$, while $\mathbf{W}=\diag(w_{i}), i=1,\dots,n_x$ is the loadings matrix. 
}

\textcolor{black}{
By definition of \textcolor{black}{left} eigenvector, $\mathbf{V}\mathbf{\Sigma_{x|y}}\mathbf{\Sigma_{x}^{-1}}=\mathbf{D}\mathbf{V}$, where $\mathbf{D}=\diag(\lambda_{i}), i=1,\dots,n_{x}$. Substituting in eq \eqref{eq:nulling_derivative} we obtain the following derivations
\begin{equation*}
    \begin{aligned}
        &[\mathbf{WDV\Sigma_{x}}\mathbf{V}^t\mathbf{W}^t+\mathbf{\Sigma_{\epsilon}}][\mathbf{WV\Sigma_{x}}\mathbf{V}^{t}\mathbf{W}^{t}+\mathbf{\Sigma_{\epsilon}}]^{-1}\mathbf{W}=\mathbf{WD}\\
        &[\mathbf{WDR}\mathbf{W}^t+\mathbf{\Sigma_{\epsilon}}][\mathbf{WR}\mathbf{W}^{t}+\mathbf{\Sigma_{\epsilon}}]^{-1}\mathbf{W}=\mathbf{WD},
    \end{aligned}
\end{equation*}
}
\textcolor{black}{
where $\mathbf{R}=\mathbf{V\Sigma_xV^t}$. Now, pre-multiplying by $\mathbf{W}^{-1}$, post-multiplying by $\mathbf{W^{-1}}(\mathbf{WRW^t}+\mathbf{\Sigma_{\epsilon}})\mathbf{W}$ and re-arranging, we end up with the following equation
\begin{equation}
\label{eq:final_derivation}
\mathbf{W^{-1}\mathbf{\Sigma_{\epsilon}}W}=\mathbf{DW^{-1}\Sigma_{\epsilon}\mathbf{W}}.
\end{equation}
}
\textcolor{black}{
Assuming that $\mathbf{D}$ is not the identity matrix (which would imply that $\mathbf{x}$ and  $\mathbf{y}$ are independent), eq. \eqref{eq:final_derivation} is verified if and only if $\mathbf{\Sigma_{\epsilon}=0}$, for any possible loading matrix $\mathbf{W}$, \textcolor{black}{which does not influence the optimization.} 
}
\bibliographystyle{IEEEtran}
\bibliography{bibliography}
\end{document}